\documentstyle[aps]{revtex}
\begin{document}
\draft
\title{Special-relativistic model flows of viscous fluid}
\author{A. D. Rogava}
\address{Department of Theoretical Astrophysics,  Abastumani  Astrophysical
Observatory, Kazbegi str. $N.~2^{a}$, Tbilisi 380060, Republic of Georgia
and Department of Physics, Tbilisi State University,
Chavchavadze ave. 2, 380028 Tbilisi, Republic of Georgia}
\date{\today}
\maketitle
\begin{abstract}
Two, the most simple cases of special-relativistic flows of a viscous,
incompressible fluid are considered: plane Couette flow and plane Poiseuille
flow. Considering only the regular motion of the fluid we found the
distribution of velocity in the fluid (velocity profiles) and the friction
force, acting on immovable wall. The results are expressed through simple
analytical functions for the Couette flow, while for the Poiseiulle
flow they are expressed by higher transcendental functions (Jacobi's elliptic
functions).
\end{abstract}


\section{Introduction}

A recent development of theoretical astrophysics revealed a wide class of
hydrodynamic and hydromagnetic flows having the close relevance to a
number of astrophysical objects such as stellar winds \cite{Ken}, accretion
flows \cite{Str}, outflows of matter in Active Galactic Nuclei
(jets) \cite{Bri} etc. The spatial symmetry of
these flows is different, thermodynamic state is sometimes unusual and
velocities are often relativistic. Relativity definitely influences the
physical processes that take place in such flows and leads to various
outstanding appearances of these objects.

The subject of the present paper is to study how the relativistic velocities
affect physical conditions and dynamics of the most simple model hydrodynamic
flows. For the present consideration we chose such examples of flows, which
being simple enough are, at the same time, relevant to the most astrophysical
cases due to their simple geometry.

In particular, we are going to study the special-relativistic generalization
of a couple of examples of the well-known classic hydrodynamic
flows {\cite{hyd}}: plane
Couette flow and plane Poiseuille flow. We consider only regular motion
in these flows. In both cases we find {\it exact} solutions of the problem.

In particular, for the plane Couette flow we  determined the velocity
distribution throughout the flow. The velocity profile is expressed
analytically and appears to be nonlinear. The nonlinearity noticeably
increases with the increase of the velocity of the moving plane and becomes
almost "steplike" for $v_0{\simeq}1$. We have found the average velocity of
the
flow that in nonrelativistic limit tends to $v_0/2$ (as it should be), while
for $v_0{\approx}1$ it tends to unity. We have also determined the friction
force acting on the immovable wall, which is exactly equal to the corresponding
classic (nonrelativistic) value.

For the plane Poiseiulle flow the velocity distribution was derived by
the expressing of the solution of the corresponding differential equation
through the
higher transcendental functions. In particular, the velocity profile is
expressed by Jacobi's elliptical functions. The general property of this
solution is that with the increase of the pressure gradient value (the very
gradient that induces the whole flow) the profile becomes more and more
"flattened."

In the concluding section we discuss the obtained results in the context
of their possible relevance to the future modeling of real relativistic
astrophysical flows.

\section{Main Consideration}

Henceforth, we shall use the following notations: greek indices will represent
spacetime components, while latin ones will be used for spatial components.
We shall use geometrical units, so that $G=c=1$. The signature of the
spacetime (Minkowskian) metric:
$$
ds^2=-d{\tau}^2=-dt^2+d{x}^2+d{y}^2+d{z}^2, \eqno(2.1)
$$
is chosen as $(-~+~+~+)$.

We shall consider motion of the incompressible, viscous fluid with the
following stress-energy tensor {\cite{Lit}}:
$$
T^{{\alpha}{\beta}}=eU^{\alpha}U^{\beta}+Ph^{{\alpha}{\beta}}-
2{\eta}{\sigma}^{{\alpha}{\beta}}-{\xi}{\theta}h^{{\alpha}{\beta}}, \eqno(2.2)
$$
where all notations are standard {\cite{Lit}}. In particular, $e$ is the
density of the mass-energy, $P$---the pressure,
$U^{\alpha}$ is the 4-velocity field of the
flow, $h^{{\alpha}{\beta}}$ is the projection tensor, defined as:
$$
h^{{\alpha}{\beta}}{\equiv}{\eta}^{{\alpha}{\beta}}
+U^{\alpha}U^{\beta}, \eqno(2.3a)
$$
$\eta$ and $\xi$ are the coefficients of the shear and bulk viscosity
respectively. $\theta$ is the "expansion" of the fluid world lines:
$$
\theta{\equiv}U^{\alpha}_{;\alpha}, \eqno(2.3b)
$$
and ${\sigma}^{{\alpha}{\beta}}$ is the shear tensor:
$$
{\sigma}_{{\alpha}{\beta}}{\equiv}{1 \over 2}{\left[U_{{\alpha};{\mu}}
h^{\mu}_{\beta}+U_{{\beta};{\mu}}h^{\mu}_{\alpha} \right]}-{1 \over 3}{\theta}
h_{{\alpha}{\beta}}. \eqno(2.3c)
$$

For the fluid with nonrelativistic temperature, density of the mass-energy
$e$ is equal to the "usual" density of matter ${\rho}$ in the proper frame
of reference.
Since a fluid is thought to be incompressible: $\rho=const$ and the continuity
equation: $({\rho}U^{\alpha})_{;{\alpha}}=0$, clearly leads to $\theta=0$
("expansion" of the incompressible fluid is equal to zero).

Under mentioned circumstances the stress-energy conservation law may be
generally written as:
$$
{T_{\alpha}}^{\beta}_{;\beta}=0. \eqno(2.4)
$$

\subsection{Plane Couette flow}

Let us consider the plane flow of the incompressible viscous fluid, which is
situated between two infinite parallel planes one of which is moving with
arbitrary velocity $v_0$. Let an $X$ axis be parallel to the planes and an $Y$
axis be normal to them so that $y=0$ in the plane at rest, while $y=L$ for
the plane
in motion. The symmetry of the problem implies that the sole nonzero component
of the fluid velocity vector field $\vec v({\vec r},t)$ is $v_x=const(x)$.
Naturally, the 4-velocity nonzero components should be equal to:
$$
U^t=(1-v_x^2)^{-1/2}{\equiv}{\gamma}, \eqno(2.5a)
$$
$$
U^x={\gamma}v_x. \eqno(2.5b)
$$
Taking into account (2.5) one easily finds that the x-th, nonzero component
of the stress-energy conservation equation (2.4) reduces to:
$$
{{{\partial}^2U_x}\over{{\partial}y^2}}=0, \eqno(2.6)
$$
and its general solution, evidently, is: $U_x=Ay+B$
where $A$ and $B$ are some constants of integration. Since the fluid at the
immovable plane ($y=0$) should be at rest, $B=0$ and then from (2.5b) we
can obtain expression for the x-th component of the 3-velocity, which is:
$$
v_x={{Ay}\over{\sqrt{1+A^2y^2}}}. \eqno(2.7)
$$

Another boundary condition: $v_x(y=L)=v_0$ helps to find the remaining
integration constant $A$:
$$
A={\gamma}_0v_0/L, \eqno(2.8)
$$
and (2.7) may be written explicitly as:
$$
v_x(y)={{v_{0}y}\over{\sqrt{L^2(1-v_0^2)+v_0^2y^2}}}, \eqno(2.9)
$$
as for ${\gamma}(y)$ we get:
$$
{\gamma}(y)=\sqrt{1+{\left({{v_0{\gamma}_0y}\over{L}}\right)}^2}. \eqno(2.10)
$$

Now, let us calculate an average velocity of this flow $\bar v$. According to
the definition we should have:
$$
{\bar v}{\equiv}{1 \over L}{\int}^{L}_{0}v_x(y)dy
$$
Taking into account (2.9) and performing necessary integration we get the
following result:
$$
{\bar v}={1 \over v_0}-{{\sqrt{1-v_0^2}}\over{v_0}}. \eqno(2.11)
$$
When $v_0{\ll}1$ (2.11) leads to the self-evident asymptotic result
${\bar v}{\simeq}v_0/2$ \cite{hyd}. For the case $v_0{\to}1$ the average
velocity ${\bar v}{\to}1$ as it, certainly, should be.

It must be noted that the tangential friction force, acting on the
immovable ($y=0$) plane, defined as:
$$
f{\equiv}2{\eta}|{\sigma}_{xy}|={\eta}|{{\partial}U_x}/{{\partial}y}|
={\eta}v_0/L, \eqno(2.12)
$$
coincides with the classic, nonrelativistic result \cite{hyd}.

\subsection{Plane Poiseuille flow}

Let us consider, now, a flow of the viscous, incompressible fluid between two
parallel infinite immovable planes, induced by the existence of the pressure
gradient along the flow axis. Let $X$ axis be coincided with the symmetry
axis of the flow. Therefore, the planes are situated at $y={\pm}L$. The
symmetry of the problem implies that:
$$
U^{\alpha}={\left[U^t, U^x{\equiv}U=f(y), U^y=0\right]}. \eqno(2.13)
$$

Note that the y-th component of the stress-energy conservation equation leads
to $P=const(y)$. The x-th component of the same equation reduces to the
following expression:
$$
(1+U^2){{{\partial}P}\over{{\partial}x}}-{\eta}{{{\partial}^2U}\over
{{\partial}y^2}}=0. \eqno(2.14)
$$

As far as $U=const(x)$ the gradient of the pressure along the flow should be
constant: ${\partial}P/{\partial}x=const$. Furthermore, this constant should
be negative if $U$ is taken to be positive. Expressing the second derivative
appearing in (2.14) as:
$$
{{{\partial}^2U}\over{{\partial}y^2}}={{{\partial}U}\over{{\partial}y}}
{{{\partial}}\over{{\partial}U}}{\left({{{\partial}U}
\over{{\partial}y}}\right)}, \eqno(2.15)
$$
we can rewrite the equation as:
$$
{1\over 2}{{{\partial}}\over{{\partial}U}}{\left[{\left({{{\partial}U}
\over{{\partial}y}}\right)^2}\right]}=-{1\over{\eta}}{\left |{{{\partial}P}
\over{{\partial}x}}\right |}(1+U^2). \eqno(2.16)
$$

Introducing new dimensionless variable: $\varepsilon{\equiv}
(2|{\partial}P/{\partial}x|/3{\eta})^{1/2}y$ we reduce (2.16) to the
following first order differential equation:
$$
{{{\partial}U}\over{{\partial}{\varepsilon}}}={\sqrt{-(U^3+3U+const)}},
\eqno(2.17a)
$$
where $const$ is some constant of integration. It may be
determined through the obvious condition, arising from the flow symmetry:
$({\partial}U/{\partial}{\varepsilon})_{{\varepsilon}=0}=0$ and
$U({\varepsilon}=0){\equiv}U_{max}$. These conditions together imply that
$const=-{U_{max}}^3-3U_{max}$. Thus (2.17a) may also be written as:
$$
{{{\partial}U}\over{{\partial}{\varepsilon}}}={\sqrt{(U_{max}-U)
(U_{max}^2+U_{max}U+U^2+3)}}, \eqno(2.17b)
$$

This equation is solved in higher transcendental functions---Jacobi's
elliptical functions (see, for details \cite{Abr} and Appendix of this paper).
Taking into account that according to (A.5):
$$
-{\lambda}{\varepsilon}=F({\varphi}{\setminus}m),
$$
where
$$
{\lambda}^2=\sqrt{3(1+U_{max}^2)}, \eqno(2.20a)
$$
$$
m={1\over2}{\left[1+{{\sqrt{3}}\over{2}}V_max\right]}, \eqno(2.20b)
$$
$$
cos{\varphi}={{{\lambda}^2-(U_{max}-U)}\over{{{\lambda}^2+(U_{max}-U)}}},
\eqno(2.20c)
$$

for $U(\varepsilon)$ we find the following expression:
$$
U(\varepsilon)=U_{max}-{\left[{{{\lambda}sn({\lambda}{\varepsilon})}
\over{1+cn({\lambda}{\varepsilon})}}\right]}^2. \eqno(2.21)
$$

As for the x-th component of 3-velocity $V{\equiv}v_x(y)$ we can simply
derive it through (2.20) via the obvious relation:
$$
V(\varepsilon)={{U(\varepsilon)}\over{\sqrt{1+U^2(\varepsilon)}}}.
\eqno(2.22)
$$

We see that with increasing of
$V_{max}$ velocity profile becomes more and more "flattened out". The similar
tendency is clearly seen also for Couette flow, considered in the previous
section.

\section{Conclusion}

In this study we consider two kinds of the special-relativistic flows of
incompressible,
viscous fluid. In particular, we examine the most simple two-dimensional
(plane) model flows between two parallel infinite planes: Couette flow
and Poiseuille flow. In both cases we find distribution of the velocity
throughout the flow (velocity profiles) and in the former case we have also
found the average velocity of the flow and the friction force, acting on
the immovable wall. The obtained solutions are {\it exact}: for Couette flow
it is analytic, while for the Poiseuille flow it is expressed by Jacobi's
elliptic functions.

As the further step of the investigation one may consider a study of
hydrodynamic instability of these flows. It is
presumable that the relativity may introduce new qualitative features in
the instability dynamics as well. However, such a study
is beyond the scope of the present paper and will be considered elsewhere.

Since the geometrical properties of considered model flows are the most
simple we think that the results obtained in the present study and the
mathematical methods of research that we are here developing, may be
relevant and useful in some kinds of prototype astrophysical relativistic
flows with comparable properties. Here we mean the relativistic regions
of accretion discs of compact objects, accretion flows and stellar winds
with spherical or quasispherical symmetry, outflows of matter in
Active Galactic Nuclei, etc. Certainly, in some of these {\it real}
astrophysical flows {\it special-relativistic} effects will be intricately
interlaced with {\it general-relativistic} effects evoked by strong
gravitational fields and {\it electromagnetic} effects (for plasma flows)
arising
due to the existence of superstrong magnetic fields. We hope that the
present study may be of some use as the standing point in the future
investigations of these complicated types of relativistic astrophysical flows.

\section{Acknowledgements}

My research was supported, in part, by International Science
Foundation (ISF) long-term research grant RVO 300.

\section{Appendix}

In the general theory of Jacobi's elliptical functions it is known
\cite{Abr} that the integral:
$$
{\int}{{dt}\over{\sqrt{-P(t)}}}, \eqno(A.1)
$$
(where $P(t){\equiv}t^3+a_1t^2+a_2t+a_3$
is the third order polynomial of $t$, having only one real root $t=\beta$)
may be reduced to the Jacobi's elliptical integral of the first kind. For
this purpose one must introduce the following coefficients:
$$
{\lambda}^2{\equiv}\sqrt{P^{\prime}(\beta)}, \eqno(A.3a)
$$
$$
m{\equiv}sin^2{\alpha}={1\over2}-{1\over8}{{{P^{{\prime}{\prime}}(\beta)}}
\over{\sqrt{P{\prime}(\beta)}}}, \eqno(A.3b)
$$
and use, also, the following substitution:
$$
cos{\varphi}={{{\lambda}^2-(\beta-t)}\over{{{\lambda}^2+(\beta-t)}}}.
\eqno(A.4)
$$
In these notations the initial integral may be written as:
$$
{\int}{{dt}\over{\sqrt{-P(t)}}}=-{1\over{\lambda}}{\int}{{d{\theta}}
\over{\sqrt{1-msin^2{\theta}}}}. \eqno(A.5)
$$

The latter integral is equal to the Jacobi's elliptical integral of the
first kind: $F({\varphi}{\setminus}m)$.

In our case $a_1=3; a_2=0$ and $a_3=const$. Despite that $a_3$ is an arbitrary
parameter, the sole real root of the polynomial appearing in (2.17) should be the
value of $U(\varepsilon)$ on the flow axis, since the symmetry of the flow
implies that at $\varepsilon=0$, ${\partial}U/{\partial}{\varepsilon}=0$ and
at this point $U=U_{max}$.

\vskip 1cm
Author's E-mail address is: {\bf andro@dtapha.kheta.georgia.su}
\end{document}